\documentclass[twocolumn,10pt,showpacs,floatfix,superscriptaddress,amsmath,amssymb,nofootinbib,aps,prd]{revtex4-1}
\usepackage{graphicx}
\usepackage{color} 

\newcommand{\ba}{\begin{eqnarray}}
\newcommand{\ea}{\end{eqnarray}}
\newcommand{\nn}{\nonumber}

\def\be{\begin{equation}}
\def\ee{\end{equation}}

\newcommand{\ul}[1]{\underline{#1}}

\begin{document}

\title{Low energy electron-phonon effective action from symmetry analysis}

\author{D.C.\ Cabra}

\affiliation{Instituto de F\'\i sica de La Plata (IFLP), CONICET \&
Departamento de F\'{\i}sica, Universidad Nacional de la Plata,\\
C.C.\ 67, (1900) La Plata, Argentina.}

\author{N.E.\ Grandi}

\affiliation{Insituto de F\'\i sica de La Plata (IFLP), CONICET \&
Departamento de F\'{\i}sica, Universidad Nacional de la Plata,\\
C.C.\ 67, (1900) La Plata, Argentina.}

\affiliation{Abdus Salam International Centre for Theoretical Physics (ICTP), Associate Scheme, \\ Strada Costiera 11, 34151, Trieste, Italy.}

\author{G.A.\ Silva}
\affiliation{Insituto de F\'\i sica de La Plata (IFLP), CONICET \&
Departamento de F\'{\i}sica, Universidad Nacional de la Plata,\\
C.C.\ 67, (1900) La Plata, Argentina}
\affiliation{Abdus Salam International Centre for Theoretical Physics (ICTP), Associate Scheme, \\ Strada Costiera 11, 34151, Trieste, Italy.}

\author{M.B.\ Sturla.}
\affiliation{Instituto de Ciencia de Materiales de Madrid (CSIC),\\ Cantoblanco, 28049, Madrid, Spain.}

\date{\today}

\begin{abstract}
Based on a detailed symmetry analysis, we state the general rules to
build up the effective low energy field theory describing a system
of electrons weakly interacting with the lattice degrees of freedom.
The basic elements in our construction are what we call the ``memory tensors'', that
keep track of the microscopic discrete symmetries into the coarse-grained action.
The present approach can be applied to lattice systems in arbitrary dimensions
and in a systematic way to any desired order in derivatives.
We apply the method to the honeycomb lattice and re-obtain the by now well-known effective
action of Dirac fermions coupled to fictitious gauge fields. As a second
example, we derive the effective action for electrons in the kagom\'e
lattice, where our approach allows to obtain
in a simple way the low energy electron-phonon coupling terms.
\end{abstract}

\maketitle

\section{Introduction}

The synthesization of graphene has triggered an enormous interest in the analysis of its low energy properties using techniques
from relativistic field theory due to the fact that electrons in this material show a Dirac like behavior for stoichiometric
fillings \cite{review}. The origin of this lies in the band structure of fermions in the honeycomb lattice: the Fermi surface degenerates at a number of points, known as ``Dirac points'', close to which the dispersion relation takes a conical shape.

The conical shape of the Fermi surface is not exclusive of graphene, but it shows up in a number of unrelated cases: fermions on the kagom\'e
lattice close to $1/3$ filling (or $2/3$ depending on the sign of the hopping
amplitude) \cite{Kagome,hyper} and fermions in the $\pi$-flux state on
the square lattice \cite{pifluxsquare} are just two examples of a long
list. One also encounters a degenerate Fermi surface in the mean field
description of strongly correlated systems for particular solutions of the
self consistent mean field equations, or in the Schwinger fermions mean field theory of
Heisenberg models \cite{Wen}. Last but not least, graphyne arises as another
candidate to show up this kind physics \cite{MNVG,KC}.
In ref.[\onlinecite{AH}] the appearance of Dirac cones on the Fermi surface was
investigated in a general way,  while in [\onlinecite{Manes11}] their
existence was  exhaustively studied for all possible non-magnetic $3$-dimensional crystals. We note in passing that Dirac cones were observed at interfaces of topological insulators \cite{FKM,TM}.

In the study of graphene, different approaches have been employed to construct the low energy effective dynamics for fermions in the presence of phonons. On the one hand, one may start from the microscopic model and follow, for the case in which the lattice is deformed, the same steps that lead to the Dirac like Hamiltonian in the undeformed lattice \cite{AS,KCN}. We call this the tight-binding  approach, see [\onlinecite{review}] for a detailed review and an extensive list of references. On the other hand, one may promote the configuration space in which the Dirac fermion propagates into a curved manifold, and then use the standard coupling of relativistic fermions to curved backgrounds\cite{Gonzales1,Gonzales2}. We call this point of view the covariant approach, see the review [\onlinecite{VozmedianoR}] and references therein. This last approach allows the description of topological defects \cite{Cortijo} as well as smooth deformed manifolds \cite{deJuan07,deJuan12}. Despite the strong bases of both
approaches, there are  some apparent discrepancies between them, a step towards solving which can be found in [\onlinecite{deJuan12}].
See also refs. [\onlinecite{iorio}] for related work applying the covariant approach.

A crucial ingredient for the construction of the low energy effective action is the symmetry group of the underlying microscopic theory. A symmetry based approach has been applied to graphene  has been proposed in \cite{Manes07} and used in several other works \cite{Winkler,basko,Falko_1,Falko_2}. The symmetries of a lattice system are given by the discrete translations of the Bravais lattice, the discrete rotations and reflections known as point symmetries, and possible internal symmetries such as spin or color. In the process of taking the continuum limit to describe the physics at sufficiently low energies, one may naively expect that any fingerprint of the underlying discrete space would be washed out, leaving a quantum field theory with no detailed information of the particular system at hand. In other words, one may expect the symmetries in the continuum limit to be given by continuous rotations and translations, together with the internal symmetries.

In the present paper we revisit in detail how symmetries are used to constrain the low energy dynamics of a generic lattice system, under the assumption that one needs to retain specific information about the underlying discreteness of the lattice symmetries. The main idea is that in the infrared, the discrete translations  of the microscopic Bravais lattice become continuous translations, while the point group retains its discrete nature. In other words, the low energy dynamics will be described by a field theory with continuous fields transforming in representations of the discrete point group. Since any term in the Lagrangian should be invariant under the full symmetry group, the system recalls its discrete origin through the invariant tensors of the point group that are needed in order to contract the fields indices. We call such invariant tensors the ``memory tensors" of the lattice. In this way, a low energy description of a system microscopically defined by certain degrees of freedom in the square lattice (whose
point group is $\sf D_4$) or in the hexagonal lattice (whose point group is $\sf D_6$) will differ in the content of memory tensors and this would lead to different invariant local terms in the effective Lagrangian.

The paper is organized as follows: In Section \ref{methodology} we give a general description of the methodology, in Section \ref{D6} we show the method at work in the case of the $\sf D_6$ point group, and we apply it to analyze both graphene and kagom\'e systems. We study as a first
example the graphene case, and show how to derive the low energy effective field
theory reproducing the tight-binding expressions \cite{AS,KCN}. These results
have also been derived in [\onlinecite{Manes07}] following a related symmetry based analysis.
The advantage of the present approach is that it can be applied to study more general cases in a straightforward manner.
In a second step, we apply the method to fermions in the kagom\'e lattice, which describes a tight-binding model at filling $1/3$ (or $2/3$, depending
on the sign of the hopping amplitude) \cite{Kagome,hyper}. In this example, the method shows
its power since the derivation of the effective action is quite straightforward. Section \ref{conclusions} contains
the conclusions and outlook and in the Appendix we present an alternative way to derive the transformation properties
of the low energy fields.
\section{General construction}
\label{methodology}
In this section we present the procedure to construct the low energy effective action of an arbitrary $d$-dimensional lattice system. We give the guidelines to build invariants out of the fields representing the low energy degrees of freedom and the memory tensors that keep track of the discrete point group symmetry in the infrared description, and discuss the basic elements defining the field theory for electrons and phonons on a lattice.
\subsection{Point group field theory}
A $d$-dimensional lattice system is defined by a unit cell containing
$A$ atoms or molecules at positions $\{\vec r_a\}_{a\in[1\cdots A]}$
within the cell,
which is repeated periodically on a Bravais lattice,
consisting in linear combinations of a set of $d$ basis vectors
$\{\vec a_i\}_{i\in[1\cdots d]}$ with integer coefficients,
$\vec R = \sum_{i=1}^d n_i\vec a_i$ with $n_i\in{\mathbb Z}$. In other words, such a system has
a configuration space consisting on sites
\be
\vec x_{\vec Ra} \equiv \vec R + \vec r_a\,,
\ee
where the components of the vectors $\vec r_a$ in the $\{\vec a_i\}_{i\in[1\cdots d]}$ basis are smaller than one and positive.
For the kind of lattices we are interested in\footnote{We concentrate on ``symmorphic'' space groups, which correspond to the case where the subset of discrete rotations and reflections is a subgroup of the space group. In the non-symmorphic case, an additional translation $\vec u_P$ that depends on the transformation $P$ should be added to the right hand side of eq. \eqref{eq:space_operator} in order to map the lattice into itself  \cite{corn}.},
a space symmetry transformation acts on the lattice sites $\vec x_{\vec Ra}$ as
\be
\vec x'_{\vec Ra} \equiv P\cdot \vec x_{\vec Ra} + \vec t\,,
\label{eq:space_operator}
\ee
where $\vec t$ is an arbitrary Bravais lattice vector, and $P$ belongs to the point group  $G_P$  which is a finite subgroup of the $d$-dimensional orthogonal group $O(d)$, consisting of discrete rotations and reflections. The group of space symmetries can be decomposed as $G_T\rtimes G_P$, where $G_T$ is the group of discrete translations of the Bravais lattice.
\subsubsection{Low energy symmetries and field theory}
In the continuum limit, the Bravais sites $\vec R$ get replaced by continuum Cartesian coordinates $\vec x$ in a $d$-dimensional Euclidean space ${\mathbb R}^d$.
In other words, the group of translations $G_T$ is replaced by
continuous translations ${\mathbb R}^d$. This implies that the
low energy degrees of freedom are well described by fields
$\Phi(\vec x)$ which are functions on ${\mathbb R}^d$. Naively one would expect that the point group would be replaced by the
continuous Euclidean point group $O(d)$, resulting in a space
symmetry group ${\mathbb R}^d\rtimes  O(d)$. If this expectation
were realized, the fields would transform under $O(d)$ and any
memory of the ultraviolet discreteness would be lost in the
infrared\footnote{In these considerations, the group $O(d)$ should
be replaced by $SO(d)$ if the low energy dynamics is not parity
invariant.}.

The point of view taken in this paper is that the last step in the above
reasoning must be omitted, the discrete group $G_P$ must be
maintained in the low energy limit as the group of point transformations.
In other words, our guiding principle is that the low energy
space symmetry group of the theory is ${\mathbb R}^d\rtimes G_P$. The
low energy degrees of freedom will therefore be described by fields
$\Phi_\ell(\vec x)$ with $\ell\in [1\cdots L]$,
transforming under a point group transformation $P\in G_P$ according to an $L$ dimensional representation of $G_P$
\be
\Phi'_{\ell}(\vec x)=\sum_{\ul\ell}[D(P)]_{\ell\ul{\ell}}\Phi_{\ul\ell}(P^{-1}\!\!\cdot\vec x)\,.
\label{fieldd}
\ee

The low energy dynamics of the system has to be invariant under ${\mathbb R}^d\rtimes G_P$. Invariance under translations is guaranteed if any term in the Lagrangian is written as an integral of local products of the fields $\Phi_\ell(\vec x)$ and their derivatives.
In order to construct invariants under $G_P$ we need to contract the $\ell$ indices with suitable $G_P$ invariant tensors, which we shall call ``memory tensors''.
A generic $G_P$-invariant tensor of order $p$ satisfies
\small
\be
T_{\ell_1\ell_2\cdots\ell_p}=\!\!\!\sum_{\ul \ell_1\ul  \ell_2\cdots\ul \ell_p}\!\!\![D(P)]_{\ul \ell_1\ell_1}[D(P)]_{\ul \ell_2\ell_2}\cdots [D(P)]_{\ul \ell_p\ell_p} T_{\ul \ell_1\ul \ell_2\cdots\ul \ell_p}\,.
\label{invariant}
\ee
\normalsize
Thus a generic   ${\mathbb R}^d\rtimes G_P$ invariant term in the Lagrangian density, of order $p$ in the fields and
containing no derivatives, reads
\be
{\cal L}_{\partial^0}=
\sum_{\ell_1 \ell_2\cdots \ell_p} T_{ \ell_1 \ell_2\cdots \ell_p}
\Phi_{ \ell_1}(\vec x)\Phi_{ \ell_2}(\vec x)\cdots \Phi_{ \ell_p}(\vec x)\,.
\label{ordenp}
\ee
Notice that derivatives $\partial_i$ transform under the vector representation of $G_P$, that in what follows we will denote as $\bf  V$. The vector representation
is defined by
\be
[D^{\bf  V}(P)]_{ij}= [P]_{ij}\,.
\label{repvect}
\ee
where $[P]_{ij}$ is the standard $d$-dimensional matrix representation of $P$.
In consequence, terms containing derivatives can be built in a similar manner, with the help of invariant tensors with as many additional vector indices as derivatives are present. Explicitly, a term of degree $q$ in derivatives and order $p$ in the fields reads
\be
{\cal L}_{\partial^q}=
\sum_{ \ell_1\cdots \ell_p, i_1\cdots  i_q} \!\!\!\!\!\!T_{ i_1 i_2\cdots  i_q \ell_1\cdots \ell_p}
\partial_{ i_1\cdots}\Phi_{\ell_1}(\vec x)\cdots \partial_{\cdots i_q}\Phi_{\ell_p}(\vec x)\,,
\label{ordenpder}
\ee
where the tensor $T_{i_1 \cdots i_q\ell_1\cdots\ell_p}$ satisfies \eqref{invariant} with the $i$ indices transforming in the vector representation.\footnote{If internal symmetries (such as spin or color) are present, the fields have additional indices transforming in a representation of the internal group, that must be contracted with the corresponding invariant tensors.}
\subsubsection{Obtention of the invariant tensors}
The $L$-dimensional representation $D(P)$ carried by the fields $\Phi_\ell(\vec x)$ is in general reducible. A reducible representation $D(P)$ can be decomposed as a sum of irreducible representations of $G_P$ as
\ba
D(P)&=& \bigoplus_{{\bf  J}} a_{\bf  J}
D^{\bf  J}(P)\,,
\label{decompos}
\ea
here $\bf  J$ labels the irreducible representations of $G_P$ and the
integer $a_{\bf  J}$ accounts for the multiplicity of the irreducible representation $\bf  J$ in the decomposition. Among the irreducible representations of $G_P$ that will be of interest in what follows we have the singlet representation $\bf  E$ in which any point group transformation acts as the identity, and the vector representation ${\bf  V}$ that we defined above \eqref{repvect}.

A given term in the action contains a product of fields and their derivatives, that transform in the corresponding tensor product representation of $G_P$. The existence of an invariant tensor to contract all the indices in such a product can be traced back to the presence of the singlet representation in the corresponding decomposition \eqref{decompos}.

To obtain the decomposition of a given representation $D(P)$, one needs to compute the character vector $\chi^D=({\rm tr}(D(P_1)),{\rm tr}(D(P_2)),\cdots)$, where $P_1$, $P_2$ are representatives of the conjugacy classes ${\cal C}_1$, ${\cal C}_2$, the dots standing for the remaining classes of the point group $G_P$. The multiplicities $a_{\bf  J}$ are obtained by linearly decomposing $\chi^D$ in terms of the characters of the irreducible representations
\be
\chi^D = \sum _{\bf  J}a_{\bf  J}\chi^{\bf  J}\,.
\label{decomp}
\ee

Given a representation $D(P)$, one finds its projection onto the irreducible representation $\bf  J$  by use of the projector \cite{stone}
\be
{\cal P}^{\bf  J}=\frac {{\rm dim}\, J}{|G_P|}\sum_{P\in G_P}
{\rm tr}\!\left(D^{\bf  J}(P)\right)^{\!*}
D(P)\,.
\ee
In order to construct an invariant tensor as those entering into eqs. \eqref{ordenp} and \eqref{ordenpder}, one needs to project  the  tensor  product representation into   its singlet component, using
\be
{\cal P}^{\bf E}=\frac {1}{|G_P|}\sum_{P\in G_P}D(P)\,.
\ee
The action of this projector is easy to understand: acting on an object that transforms under $D(P)$, it gives the sum of all its images under point group transformations. The resulting object will necessarily be invariant under $G_P$. Explicit examples will be given below.
\subsection{Phononic and electronic fields}
The natural question is then which representations of $G_P$ have to be considered when constructing the effective field theory of a
given lattice system, or in other words, what is the correct field content $\Phi_\ell(\vec x)$. In this section we obtain the field content that takes into account the low energy degrees of freedom related to phonons and electrons moving in the lattice.
\subsubsection{Phononic field}
To analyze the vibrational degrees of freedom of the lattice we denote by $v_{I a}(\vec R)$ the displacement of the atom $a$
on the Bravais site $\vec R$ in the $I$-th direction
of $3$-dimensional space $I\in[1\cdots3]$. Notice that for the case of a lower dimensional $d<3$ lattice, the $v_{Ia}$ include the transverse displacements as well as the displacements along the directions of the lattice.
They give a total of $3A$ degrees of freedom which enter  quadratically in the Lagrangian at the harmonic level.
The resulting $3A$
normal modes can be classified, according to the standard theory of phonons in solids, as $3A-3$ optical modes plus $3$ acoustic
modes \cite{born}. At sufficiently low energy only the acoustic modes are relevant, and they have linear dispersion relation. In consequence,
at low energy we need to consider only three fields $v_I(\vec x)$.

The fields $v_I(\vec x)$ can be decomposed as $v_I = (u_1, \cdots, u_d, h_1, \cdots, h_{3-d})$. The $u_i(\vec x)$ fields take into account the deformations of the Bravais lattice
and transform  under the point group $G_P$ in the vector representation ${\bf V}$ defined in \eqref{repvect}. On the other hand, the $3-d$ fields $h_r(\vec x)$ account for the transverse (or ``flexural'') displacements, and they are scalars {\em i.e.}   transform in the singlet representation ${\bf E}$.

To isolate the elastic degrees of freedom from the global translations and rotations of the lattice on the Euclidean 3-dimensional space in which it is embedded, the effective action must depend only on the combinations $\partial_i h_r(\vec x)$ and $u_{ij}(\vec x)=1/2(\partial_i u_{j}(\vec x)+\partial_j u_{i}(\vec x))$ \cite{landau}. The field $\partial_i h_r(\vec x)$ transform in the vector representation
\be
\partial_ih_r'(\vec x) = \sum_{\ul i} P_{i\ul i}\  \partial_{\ul i}h_r(P^{-1}\!\!\cdot \vec x)\,
\ee
and we denote this as
\be
\partial_i h\in {\bf V}\,.
\ee
Regarding $u_{ij}(\vec x)$, it is a symmetric tensor that transforms in the product representation ${\bf V}\otimes {\bf V}$ as
\be
u_{ij}'(\vec x) = \sum_{\ul i\ul j} P_{i\ul i}P_{j\ul j}\  u_{\ul i\ul j}(P^{-1}\!\!\cdot \vec x)\,,
\ee
in other words
\be
u_{ij} \in{\cal S}({\bf V}\otimes{\bf V})\,.
\ee
where ${\cal S}$ stands for ``symmetric part''. Notice that any symmetric tensor can be decomposed into its trace and its symmetric traceless parts, namely
\be
u_{ij}(\vec x)=\frac1d\delta_{ij}\sum_k u_{kk}(\vec x)+\sum_{kl}\left(\delta_{ik}\delta_{jl}-\frac1d\delta_{ij}\delta_{kl}\right)u_{kl}(\vec x)\,.
\ee
Since the first term is invariant under point group transformations,  the representation ${\cal S}({\bf V}\otimes{\bf V})$ contains {\bf a} singlet in its decomposition, {\it i.e.} ${\cal S}({\bf V}\otimes{\bf V})={\bf E}\oplus\cdots$. This singlet describes the homogeneous dilation of the lattice.
\subsubsection{Electronic field}
In order to derive the transformation properties for the electronic degrees of freedom,
we need to work out the implications of eq. \eqref{eq:space_operator}. To do that, we note that under a point group transformation $P$, the position $\vec r_a$ of the $a$-th
atom inside the unit cell transforms as
\be
\vec r_a'\equiv P\cdot \vec r_a =\vec r_{a_P}+\vec t_a^{\,P}\,.
\label{eq:vector}
\ee
Then under a point group transformation the atom $a$ in the Bravais site ${\vec R}$ is mapped into the atom $a_P$ in the Bravais site $P\cdot\vec R+\vec t_a^{\,P}$. The Bravais lattice vector $\vec t_a^{\,P}$ arises whenever the action of $P$ takes $\vec r_a$ out of the unit cell (see Fig. \ref{figure}).

Calling  $c_{a}(\vec R)$ the operator that annihilates an electron in the Wannier state localized around the site $\vec x_{\vec Ra}$, a point group transformation $P$ acting on the lattice
maps such state into the corresponding Wannier state centered around the transformed point $P\cdot  \vec x_{\vec Ra}$, namely
\footnote{To prove this, let us write the Wannier wavefunction centered around the atom $a$ on the site $\vec R$ as $\psi_{\vec Ra}(\vec r)=\psi_a(\vec r - \vec R -\vec r_a)$, an electron in this state is annihilated by the operator $c_{a}(\vec R)$. Applying a point group transformation $P$ to the space coordinates $\vec r$ we get
\ba
\psi'_{\vec Ra}(\vec r)&=&\psi_{\vec Ra}(P^{-1}\cdot\vec r)
=\nn\\&=&
\psi_a(P^{-1}\cdot\vec r - \vec R -\vec r_a)
=\nn\\&=&
\psi_a(P^{-1}\cdot(\vec r - P\cdot\vec R -P\cdot\vec r_a))
=\nn\\&=&
\psi_{a_P}(\vec r - P\cdot\vec R -\vec r_{a_P}-\vec t_a^{\,P})
=\nn\\&=&
\psi_{(P\cdot\vec R +\vec t_a^{\,P})a_P}(\vec r)
\ea
where in going from the third to the fourth line, we used \eqref{eq:vector} and the fact that the function $\psi_a(\vec r)$ satisfies $\psi_{a_P}(\vec r)=\psi_a(P^{-1}\cdot\vec r)$. This implies that an electron in the transformed wavefunction will be annihilated by $c_{a_P}(P\cdot\vec R +\vec t_a^{\,P})$.}
\be
 c'_{a}(\vec R)=  c_{a_P}(P\cdot \vec R+\vec t_a^{\,P}) \,.
\label{eq:trafo_wannier}
\ee

Bloch states are annihilated by the operators
\be
c_{a}(\vec k)\equiv \sum_{\vec R\in {\rm Bravais}}e^{-i\vec k\cdot \vec R}\,c_{a}(\vec R)\,.
\ee
Using \eqref{eq:trafo_wannier}, one finds that a point group transformation $P$ acts on Bloch states as
\be
c'_{a}(\vec k)= \sum_{\vec R\in {\rm Bravais}}e^{-i(P\cdot\vec k)\cdot (\vec R-\vec t_a^{\,P})}\,
c_{{a_P}}(\vec R)\,.
\ee
In other words we obtain \cite{lomer}
\be
c'_{a}(\vec k)= e^{i(P\cdot\vec k)\cdot \vec t_a^{\,P}}
c_{{a_P}}(P\cdot\vec k) \ .
\label{lapindon}
\ee
As will become clear below, the phase factor on the right hand side characterizes the transformation properties of the
low energy electronic degrees of freedom under point group transformations.

In a non-interacting system, the Hamiltonian is diagonalized by linear combinations of Bloch states
with fixed $\vec k$.
Since the index $a$ takes $A$ values (corresponding to $A$ atoms or molecules in the unit cell), there would
generically be $A$ different energy bands $\{\varepsilon_n(\vec k)\}_{n\in[1\cdots A]}$. Calling
$c_n(\vec k)$ the operator that creates an energy eigenstate with crystalline momentum $\vec k$ in the $n$-th band, we can write
\be
c_a(\vec k)=\sum_{n=1}^A \alpha_{a n}\, c_{ n}(\vec k)\,.
\label{14}
\ee
Bloch theorem implies
that states with crystalline momentum $\vec k$ accommodate in representations of
the group $G_{\vec k}$ consisting in those point group transformations that leave $\vec k$ invariant, the so called ``little group'' of
$\vec k$
\footnote{The proof is as follows: Bloch's theorem states that the eigenfunctions of an electron in a Bravais lattice can be written as $\psi_{n\vec k}(\vec r)=\exp(i\vec k\cdot\vec r)u_{n\vec k}(\vec r)$ where the function
$u_{n\vec k}(\vec r)$ is periodic in the lattice and satisfies
\be
\left[
{\frac {\hbar^2}{2m}}
	\left(
	-\vec \nabla^2-2i\ \vec k\cdot\vec \nabla+\vec k^2
	\right)
	+V(\vec r)
	\right]u_{n\vec k}(\vec r)=\varepsilon_n(\vec k)\,u_{n\vec k}(\vec r)\,.\nn\\
\nn
\ee
This equation must be solved in the unit cell with periodic boundary conditions. Although the potential
$V(\vec r)$ is invariant under the action of the full point group $G_P$, the equation is only invariant under the restricted set of point transformations that satisfy $P\cdot\vec k=\vec k$, {\em i.e.} under the little group $G_{\vec k}$ of the momentum $\vec k$. The eigenfunctions $\psi_{n\vec k}(\vec r)$ therefore accommodate in a representation of {$G_{\vec k}$}.}.
This in turn implies that, if there exist values of $\vec k$ such that
the transformation rule  \eqref{lapindon} contains a non-trivial representation of the little group $G_{\vec k}$,
then the $n$ indices mix under little group transformations, and the energy bands will be degenerate at those points.
We will assume that such degeneracy occurs at isolated points $\{k_m\}_{m\in[1\cdots M]}$.

The ground state of the non-interacting electron system is obtained by filling all Bloch states with negative energy $\varepsilon_{n}(\vec k)<0$ while
those with positive energy $\varepsilon_{n}(\vec k)>0$ are empty. We will assume that interactions do not alter this picture.

Now we are ready to take the low energy limit. Given an energy cutoff $\varepsilon_{cutoff}$, high energy bands, those which lie completely above it, are not excited at all at low energies. On the other hand, low energy bands
can be excited only for momenta satisfying $|\varepsilon_n(\vec k)|<\varepsilon_{cutoff}$.

We will assume in what follows that low energy bands have their minima or maxima at degenerate points $\{\vec k_m\}_{m\in[1\cdots M]}$ and we will tune the chemical potential to ensure that those points lie at zero energy. In consequence low energy states have momentum close enough to some of the $\vec k_m$. To construct the low energy effective action we replace the annihilation operators by
\be
c_{nm}(\vec q)= c_{n}(\vec k_{m}+\vec q)\,.
\label{repij}
\ee
With these operators we can recover, in our original basis \eqref{14}
\be
c_{ma}(\vec q)\ \ = \ \sum_{ n}  \alpha_{an} \,c_{nm}(\vec q)\,.
\label{donga}
\ee

Notice that in \eqref{repij} we have also expanded the high energy bands annihilation operators around the minima of the low energy bands, and that we have included the resulting operators in the sum of \eqref{donga}. Since this corresponds to an assumption on the high energy modes, it would be harmless as long as the corresponding additional degrees of freedom are gapped in the effective theory. This observation will be important in Section \ref{D6} when we study the kagom\'e lattice.

Since under point group transformations the energy bands have to be invariant, then the transformation $P$ of one of the minima
$\vec k_{m}$ must necessarily result in another minimum $\vec k_{m_P}$, namely
\be
P\cdot \vec k_{m} = \vec k_{m_P}.
\label{kmp}
\ee
With this, and using formula
\eqref{lapindon}, we can state that under point group transformations low energy annihilation operators transform according to
\be
c'_{ma}(\vec q)= e^{i\vec k_{m_P}\cdot \vec t_a^{\,P}}
c_{m_Pa_P}(P\cdot\vec q)\,,
\label{lachot}
\ee
where, in the low energy limit, we have disregarded the $\vec q$ dependence of the exponent in the right hand side of \eqref{lapindon}.
By Fourier transforming this low energy operators, we get our electronic annihilation fields
\be
\Psi_{ma}(\vec x)=\int d^dq\,
e^{i\vec q\cdot\vec x}c_{ma}(\vec q)\,.
\ee
The coordinate $\vec x$ in the above equation denotes the position in the Bravais lattice.
The transformation rule \eqref{lachot} determines the representation of the point group under which the electronic fields transform,
its properties being fixed by the transformation rule of the atom positions $\vec r_a$ and of the minima $\vec k_{m}$ of the low energy bands. Explicitly
\ba
\Psi'_{ma}(\vec x)&=&  \int d^dq\,
e^{i\vec q\cdot\vec x}c'_{ma}(\vec q) \nn\\
&=& e^{i\vec k_{m_P}\cdot \vec t_a^{\,P}} \Psi_{m_Pa_P}(P\cdot \vec x)
\label{psi}
\ea
To make the above equation compatible with the general transformation rule \eqref{fieldd} of a field, we need to replace $P$ in the above formula with $P^{-1}$, to obtain\footnote{This is consistent with the fact that transformation $P^{-1}$ on the wavefunction coordinates $\vec r$ acts as a transformation $P$ on the lattice.}
%
\be
\Psi'_{ma}(\vec x)   = \sum_{\ul m\ul a}[D(P)]_{ma\ul m\ul a}  \Psi_{\ul m\ul a}(P^{-1}\cdot \vec x)\,,
\label{lepen}
\ee
with
\be
[D(P^{-1})]_{ma\ul m\ul a} = \delta_{\ul m m_P} \delta_{\ul aa_P} e^{i\vec k_{m_P}\cdot \vec t_a^{\,P}}\,,
\label{lagarch}
\ee
where the indices $m,a$ determine $m_P, a_P$.
We denote this representation as $\bf R$, {\it i.e}
\be
\Psi_{ma} \in {\bf R}\,.
\ee
Notice that this representation is in principle reducible.

\vspace{.5cm}

To summarize, in order to construct the low energy field theory we need to identify the discrete point group $G_P$.  Low energy fields representing phonons consist on flexural $\partial_i h_r(\vec x)$ and in-plane $u_{ij}(\vec x)$ degrees of freedom, transforming in the vector ${\bf V}$ and symmetric product ${\cal S}\left({\bf V}\otimes {\bf V}\right)$ representations of $G_P$ respectively, while those representing electrons consist on $\Psi_{ma}(\vec x)$ transforming in the representation $\bf R$ defined in \eqref{lepen}. In consequence our field content is given as $\Phi_\ell(\vec x)=(\partial_i h_r(\vec x), u_{ij}(\vec x),\Psi_{ma}(\vec x))$, and  invariant terms will be built out of memory tensors with indices in the corresponding representations. The list of such tensors can be exhausted by group theory methods, thus providing the most general effective action compatible with the point symmetries. Up to this point we have only made use of the point symmetry group of the system. At low energies, the
translation subgroup enters only through its continuum description, which implies the conservation of $\vec q$. Nevertheless, the underlying discrete translation symmetry imposes the conservation of crystalline momentum, thus a term in the Lagrangian originating transitions between states at different minima $\vec k_m$ in the Brillouin zone must be forbidden. This implies that, among all the possible invariant tensors, we must keep only those which are completely diagonal in $m$.

In the next section we apply this procedure to a particular important case.
\section{$\sf D_6$ field theory}
\label{D6}
In this section we apply the methodology described above to
the case in which the point group is $\sf D_6$, which arises, in
particular, for the cases of the hexagonal and kagom\'e lattices.

We first discuss the common features of these two different lattices, {\it i.e.}
the decomposition of reducible representations into irreducible ones that allows us to construct the invariant tensors.

In a second step we construct the
singlet terms ($\sf D_6$-invariants) that conform
the low energy effective action as an expansion in momenta. The results are compared with the
literature. At this points we analyze the
different lattices separately, showing how to derive the relevant invariant terms in each case.
\subsection{$\sf D_6$ group and its representations}
In general, the ``dihedral group'' ${\sf D}_n$ is defined as the symmetry group of a regular $n$-polygon.
It is generated by two elements $R,F_x$ as
\be
{\sf D}_n=\left\{ R,F_x~|~R^n=e,~F_x^2=e,~F_x^{-1}RF_x=R^{-1}\right\}\,,
\label{constraints}
\ee
here $e$ is the neutral element of the group.

The $\sf D_6$  group  is a non-abelian discrete group of order $|\sf D_6|=12$. It can
be generated from a $60^o$ rotation $R$ and a reflection through the $x$-axis $F_x$.
It has six conjugacy classes: those of the identity and $R^3$ of dimension 1, those or $R$ and $R^2$ of dimension $2$ and those of $F_x$ and $R F_x$ of dimension $3$.

For completeness we mention that $\sf D_6$ is a product group $\sf D_6=D_3 \times \mathbb Z_2$ with $\sf D_3$ the symmetry group of an equilateral triangle generated from a $120^o$ rotation and a reflection.
\subsubsection{Representations}
The $\sf D_6$ group has six irreducible representations, that we denote by
\be
{\bf J} = {\bf E}, {\bf A}_1, {\bf A}_2, {\bf A}_3, {\bf V}, {\bf V}'\,,
\ee
where $\bf E$ is the singlet representation, ${\bf A}_1,{\bf A}_2,{\bf A}_3$ are 1-dimensional and $\bf V$,${\bf V}'$ are 2-dimensional. $\bf V$ is the vector representation \eqref{repvect} and we refer to ${\bf A}_1$ and ${\bf V}'$ as ``pseudoscalar" and ``pseudovector" representations respectively.

The pseudoscalar representation ${\bf A}_1$, which will appear below, corresponds to representing the rotations
trivially and the reflections by a sign.

The (faithful) vector representation $\bf V$ is written
naturally in terms of $2\times2$ orthogonal matrices acting on the plane
\ba
&&D^{\bf V}(R)=\left(
          \begin{array}{cc}
                    \frac12 & -\frac{\sqrt3}2 \\
                    \frac{\sqrt3}2 & \frac12 \\
                  \end{array}
                \right),
\\
&&D^{\bf V}(F_x)=\left(
     \begin{array}{cc}
       1 & 0 \\
       0 & -1 \\
     \end{array}
   \right)\,.
   \label{irrepv}
\ea

The (non-faithful) pseudovector representation $\bf V'$ can be obtained from the $\bf V$
matrices as
\ba
&&D^{\bf V'}\!(R)=(D^{\bf V}(R))^2=\left(
          \begin{array}{cc}
                    -\frac12 & -\frac{\sqrt3}2 \\
                    \frac{\sqrt3}2 & -\frac12 \\
                  \end{array}
                \right),\
\label{vpr}
\\
&&D^{\bf V'}\!(F_x)=D^{\bf V}(F_x)\,=\left(
     \begin{array}{cc}
       1 & 0 \\
       0 & -1 \\
     \end{array}
   \right)\,.
   \label{vprima}
\ea
Due to non-faithfulness of $\bf V'$ one has
\ba
D^{\bf V'}\!(F_y)&=&D^{\bf V'}\!(R^3F_x)=
\nn\\
&=&(D^{\bf V'}\!(R))^3D^{\bf V'}\!(F_x)
=
\nn\\
&=&D^{\bf V'}\!(F_x)\,,
\label{parida}
\ea
where we have used \eqref{vpr} to go from the second to the third line in \eqref{parida}. Hence,  reflections
through  $x$ and $y$-axis act on a $\bf V'$ pseudovector in the same way.

In what follows, to distinguish vectors in the 2-dimensional representation $\bf V$ from pseudovectors in the 2-dimensional representation $\bf V'$, we use for the first Latin indices, while for the second we use Greek indices.

The decomposition of a given $\sf D_6$ representation $D(P)$ into irreducible representations amounts to compute the character
\be
\chi^D=({\sf tr}(e),{\sf tr} (R^3) ,{\sf tr} (R) ,{\sf tr} (R^2) ,{\sf tr} (F_x) ,{\sf tr} (RF_x) )\,,
\label{char}
\ee
where ${\sf tr}(P)={\rm tr}(D(P))$ with $P\in \sf D_6$, and to write it as a linear combination of the characters of the irreducible representations. The character table for $\sf D_6$ can be found in \cite{bradley}
\begin{center}
\begin{tabular}{|l||c|c|c|c|c|c|}
\hline
 $\sf D_6$ & ${\cal C}_e$ & ${\cal C}_{R^3}$ & $2\,{\cal C}_R$ & $2\,{\cal C}_{R^2}$ & $3\,{\cal C}_{F_x}$ & $3\,{\cal C}_{RF_x}$ \\
 \hline\hline
  $\bf E$ & 1 & 1 & 1 & 1 & 1 & 1 \\\hline
  ${\bf A}_1$ & 1 & 1 & 1 & 1 & -1 & -1 \\\hline
  ${\bf A}_2$ & 1 & -1 & -1 & 1 & -1 & 1 \\\hline
  ${\bf A}_3$ & 1 & -1 & -1 & 1 & 1 & -1 \\\hline
  $\bf V'$ & 2 & 2 & -1 & -1 & 0 & 0 \\\hline
  $\bf V$ & 2 & -2 & 1 & -1 & 0 & 0 \\\hline
 \end{tabular}
\end{center}
The hor\-i\-zontal lines of this ta\-ble contain six char\-ac\-ter vectors $\chi^{\bf J}$ one for each ir\-reducible rep\-resen\-tation. These
vec\-tors are orthogonal with the metric $g_{ij}=\mathrm{diag}(1,1,2,2,3,3)$
whose diagonal entries are given by the dimension of each of the classes (quoted in the top line of the table).

Decomposing \eqref{char} in the basis given by $\{\chi^{\bf J}\}$ as in \eqref{decomp} one obtains the coefficient
$a_{\bf J}$ giving the number of times the irreducible representation $\bf J$ appears in the representation $D(P)$.

For the construction of
invariants it is useful to quote the decomposition of the product of any
two irreducible representations. One has
\be
\bf E\otimes Irrep=Irrep \,,\nn
\ee
\vspace{-.5cm}
\be
\begin{array}{lll}
{\bf A}_1\otimes {\bf A}_1={\bf E} ~~(similar~for~{\bf A}_2,{\bf A}_3),
& {\bf A}_1\otimes\bf  V'=V'\,,
\\
{\bf A}_1\otimes {\bf A}_2={\bf A}_3\,(and~cyclics),
&
{\bf A}_1\otimes \bf V=V    \,,
\\
{\bf A}_2\otimes{\bf  V=V'},
&
{\bf A}_2\otimes {\bf V'=V} \,,
\\
{\bf A}_3\otimes{\bf  V=V'},
&
{\bf A}_3\otimes {\bf V'=V}\,,
\end{array}\nn
\ee
\ba
\begin{array}{lll}
 \bf V\otimes V&=&{\bf E}\oplus {\bf A}_1\oplus\bf V'\,, \\
 \bf V\otimes V'&=&{\bf A}_2\oplus {\bf A}_3\oplus\bf  V \,,\\
 \bf V'\otimes V'&=&{\bf E}\oplus {\bf A}_1\oplus \bf V'\,.
\end{array}
\label{CG}
\ea
These rules allow us to decompose and arbitrary product of representations.
%
%
%
%
%
%
\subsubsection{Memory tensors}
As mentioned above, the memory tensors have their origin in the decomposition of products of representations into irreducible representations. As an example, the first line in \eqref{CG}
indicates that a product of two vector representations decomposes as a sum of three irreducible representations: the first two are 1-dimensional and the third one is 2-dimensional.

The presence of $\bf E$ in the decomposition of ${\bf V}\otimes{\bf V}$ indicates the existence of a tensor with which we can construct a scalar out of two vectors. To obtain such tensor, we apply the projector ${\cal P}_{\bf E}$ to an object $o_{ij}$ transforming in the product representation, to obtain
\ba
[{\cal P}^{\bf E} o]_{ij} &=&\frac 1{12}\sum_{P}P_{\ul i i}P_{\ul j j} o_{\ul i \ul j}=\nn\\
&=& \frac12(o_{11}+o_{22})\delta_{ij}\,.
\ea
Stripping the scalar factor, we extract our invariant tensor in the ${\bf V}\otimes{\bf V}$ representation. In a similar way, one can extract the pseudoscalar part ${\bf A}_1$ as
\ba
[{\cal P}^{{\bf A}_1} o]_{ij} &=&\frac 1{12}\sum_{P}s(P)P_{\ul i i}P_{\ul j j} o_{\ul i \ul j}=\nn\\
&=& \frac12(o_{11}-o_{22})\epsilon_{ij}
\ea
where $s(P)$ is the sign representing the transformation $P$ in the pseudoscalar representation ${\bf A}_1$. The resulting tensor $\epsilon_{ij}$ gives a pseudoscalar when contracted with two vectors. This tensor can be also understand as the singlet in the decomposition
\be
{\bf A}_1\otimes{\bf V}\otimes{\bf V}={\bf E}\oplus{\bf A}_1\oplus{\bf V}'
\ee

Finally, the presence of ${\bf V}'$ in the ${\bf V}\otimes{\bf V}$ decomposition calls for the existence of a $\sf D_6$-invariant three index tensor $S_{\alpha ij}$ such that when contracted with two vectors gives a pseudovector. Its existence can also be inferred from the singlet in the decomposition
\ba
\bf V\otimes V \otimes V'&=&{\bf E}\oplus {\bf A}_1\oplus 3\bf V' \label{vvvp}\,,
\ea
Such tensor can be obtained by using the corresponding projector, and it reads
\be
S_{1 ij}=(\sigma_3)_{ij},~~~~~S_{2 ij}=(\sigma_1)_{ij}\,.
\label{nicoviri}
\ee
where $\sigma_1, \sigma_3$ are the standard Pauli matrices. According to \eqref{invariant} the invariance of the tensor $S_{\alpha ij}$ under $\sf D_6$ implies that it satisfies
\be
 S_{\alpha i j}= \sum_{\ul \alpha \ul i\ul  j}[D^{\bf V'}(P)]_{\ul \alpha\alpha}\,[D^{\bf V}(P)]_{\ul ii}\,[D^{\bf V}(P)]_{\ul jj}\,S_{\ul \alpha \ul i\ul j }\,.
\ee
This tensor turns out to be a key ingredient in the construction of the effective action. 

For completeness we observe that from the last line in \eqref{CG} we can infer the existence of a tensor $\delta_{\alpha\beta}$ that provides a scalar out of two pseudovectors, a tensor $\epsilon_{\alpha\beta}$ providing a pseudoscalar, and a three index tensor $T_{\alpha \beta\gamma}$ providing a pseudovector. This last tensor can be understood as the singlet in the decomposition
\ba
 \bf V'\otimes V'\otimes V'&=&{\bf E}\oplus {\bf A}_1\oplus 3\bf V'\,.
\label{vpvpvp}
\ea
Its explicit components are given by
\be
T_{1 \alpha\beta}=(\sigma_3)_{{\alpha\beta}},~~~~~T_{2 {\alpha\beta}}=-(\sigma_1)_{{\alpha\beta}}\,.
\label{gui}
\ee
The $T_{\alpha\beta\gamma}$ tensor \eqref{gui} is totally symmetric in its indices and traceless on any two indices. It satisfies
\be
T_{\alpha \beta\gamma}= \sum_{\ul \alpha \ul \beta\ul \gamma}[D^{\bf V'}(P)]_{\ul \alpha\alpha}\,[D^{\bf V'}(P)]_{\ul\beta \beta}\,[D^{\bf V'}(P)]_{\ul \gamma\gamma}\,T_{\ul\alpha \ul\beta\ul\gamma }\,.
\label{gtransprop}
\ee
This tensor allows to obtain the singlet in \eqref{vpvpvp} out of the product of three $\bf V'$ pseudovectors.
\subsection{From memory tensors to the effective action}
Let us first analyze the symmetry properties of the low energy degrees of freedom we want to describe, that is find the field content describing electrons and phonons on our $\sf D_6$ field theory. First we discuss the phonon modes, which have the same symmetry properties for the two examples we want to analyze in detail, {\it i.e.} the hexagonal and
the kagom\'e lattices. Then we turn into the electron modes, which have different transformation properties in each system.
\subsubsection{Phonons}
As mentioned in Section \ref{methodology}, the elastic degrees of freedom split into in-plane phonons $u_{ij}(\vec x)$ transforming in the symmetric tensor representation ${\cal S}({\bf V}\otimes {\bf V})$ and out of plane flexural phonons $\partial_i h_r(\vec x)$ transforming in the vector representation $\bf V$. In the present case, $i=x,y$ and the index $r$ can be omitted since there is a single transverse direction.

The vector representation is generated by the matrices \eqref{irrepv} and it is irreducible. Then for the flexural phonons we can simply write
\be
\partial_i h\in {\bf V}\,.
\ee

On the other hand, the tensor representation ${\bf V}\otimes {\bf V}$ decomposes as in the first line of \eqref{CG} in terms of irreducible representations. It consists of tensors with two vector indices, which allows us to use the standard decomposition of a tensor into its trace, its antisymmetric and its symmetric parts, in order to get an interpretation of the irreducible representations entering into \eqref{CG}. Indeed, as advanced in Section \ref{methodology}, the singlet representation ${\bf E}$ corresponds to the trace part. Being two dimensional, the ${\bf V}'$ representation corresponds to the two dimensional space of symmetric traceless $2\times2$ tensors. Finally, the ${\bf A}_1$ representation must then correspond to the one dimensional space of antisymmetric $2\times2$ matrices. With this, we can deduce
\be
u_{ij}\in{\cal S}({\bf V}\otimes{\bf V})={\bf E}\oplus{\bf V}'\,.
\label{pho}
\ee
We can arrange the two independent components of the symmetric traceless part of $u_{ij}(\vec x)$ in a ${\bf V}'$ pseudovector, by making use of our invariant tensor $S_{\alpha ij}$ as
\ba
A_\alpha^{(u)}(\vec x)&=&\sum_{ij}S_{\alpha ij}\,u_{ij}(\vec x)=
\nn\\
&=&(\partial_xu_x-\partial_yu_y,\partial_xu_y+\partial_yu_x)\,.
\label{Adu}
\ea
We would like to stress that $A_\alpha^{(u)}(\vec x)$ is not a vector in the usual sense since it transforms in the $\bf V'$ representation. This object has been named pseudovector since under a $y$-axis reflection  $(x,y)\to(-x,y)$ transforms as $(A_x^{(u)}(\vec x),A_y^{(u)}(\vec x))\to(A_x^{(u)}(\vec x),-A_y^{(u)}(\vec x))$ contrary to
naive expectations.
This last transformation is nothing but the expected one as soon as one recognizes that $A_\alpha^{(u)}(\vec x)$ transforms as a $\bf V'$ representation of $\sf D_6$
\be
{A_{\alpha}^{(u)}}'(\vec x)=\sum_\beta [D^{\bf V'}(P)]_{\alpha\ul\alpha}\,{A}_{\ul\alpha}^{(u)}(P^{-1}\cdot\vec x)\,,\label{yep}
\ee
where $D^{\bf V'}(P)$ are as in \eqref{vpr}, \eqref{vprima}.
\subsubsection{Electrons}
In order to obtain the representation in which the electronic fields transform, we follow the lines sketched in Section \ref{methodology}. The two examples we study, namely that of graphene and electrons on the kagom\'e lattice, correspond to a triangular lattice with symmetry group ${\sf D}_6$.
The first Brillouin zone consists of a hexagon whose vertices are identified under $120^o$ rotations. In other words, we have two independent vertices $k_m$ with $m=+,-$, which have a non-trivial little group. As explained in Section \ref{methodology}, this implies that at these points $\vec k_{\pm}$, the energy bands degenerate, and our technique applies as long as the filling selects the corresponding energy as the ground state. This is precisely the case for graphene at stoichiometric fillings and in the kagom\'e case we adjust the chemical potential to have filling $1/3$.

The triangular Bravais lattice is generated as $\vec R=n_1\vec a_1+n_2\vec a_2$ with $n_1,n_2\in{\mathbb Z}$ and
\be
\vec a_1=(1,0),  ~~~~\vec a_2=(\cos\frac\pi3,\sin\frac\pi3),
\label{bravaisvect}
\ee
where for convenience we set the lattice spacing to one by an appropriate choice of units. The two inequivalent Fermi points are located at
\be
\vec k_\pm=\pm\frac{4\pi}{\sqrt{3}}(1,0).
\ee

~

\paragraph{The graphene case}

~

~

Being made out of carbon atoms arranged in a hexagonal structure, graphene can be seen as a triangular lattice with a fundamental cell containing two atoms at positions $r_a$ with $a={\sf a},{\sf b}$. The low energy dynamics is hence described by two energy bands.
The positions of the ${\sf a}$ and ${\sf b}$ atoms inside the unit cell are
\be
\vec r_{\sf a}=\frac1{\sqrt3}(\cos\frac\pi6,\sin\frac\pi6),~~~~~\vec r_{\sf b}=\frac2{\sqrt3}(\cos\frac\pi6,\sin\frac\pi6).
\ee

\begin{figure}
\vspace{-1cm}
\includegraphics[width=0.55\textwidth]{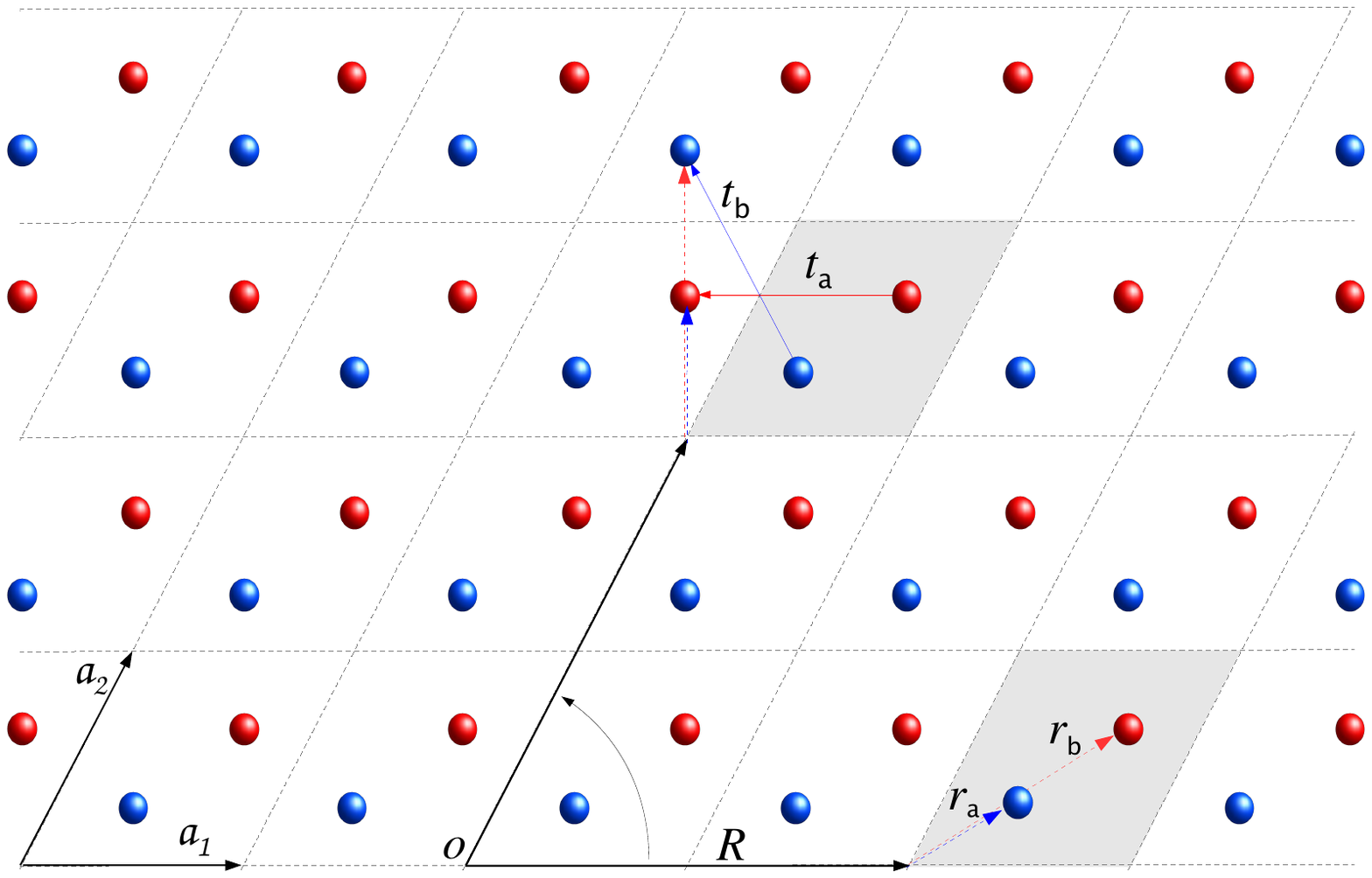}
\vspace{-8.5cm}
\caption{In the honeycomb lattice, a rotation $P$ maps the atoms $\sf a$ and $\sf b$ corresponding to the cell at position $\vec R$, into the atoms $\sf b$ and $\sf a$ corresponding to two different cells, none of which sits at the rotated position $P\cdot\vec R$. In order to express the positions of the rotated atoms in terms of the atoms corresponding to the rotated cell, we need to use the vectors $\vec t_{\sf a}$ and $\vec t_{\sf b}$.}
\label{figure}
\end{figure}

To analyze the low energy limit we consider the neighborhood of the Dirac points $\vec k_{\pm}$ and write $\vec k=\vec k_\pm+\vec q$ with $|\vec q|$ small enough. We expect that only the degrees of freedom in such regions participate in the dynamics and therefore we get four low energy independent annihilation fields
\be
\Psi_{ma}(\vec x) =\left(\begin{array}{c}
               \Psi_{+\sf a}\\
               \Psi_{+\sf b}\\
               \Psi_{-\sf a}\\
               \Psi_{-\sf b}
              \end{array}
   \right).
   \label{vac}
\ee
These fields transform according to \eqref{lepen}, \eqref{lagarch} and to make the transformation rule explicit, we need to compute $\vec t_{a}^{\,P}$, $a_P$ and $m_P$ defined in \eqref{eq:vector}, \eqref{kmp}. In order to do that, let us  first notice that the action of the {\sf D}$_6$ generator $R$ on Bravais sites is homogeneous, while for the unit cell basis $\vec r_{\sf a}, \vec r_{\sf b}$ one gets a non-homogeneous piece (see Fig. \ref{figure})
\ba
R \cdot \vec r_{\sf a} &=& \vec r_{\sf b}
- \vec a_1\,,
\nn\\
R\cdot \vec r_{\sf b} &=& \vec r_{\sf a}+
\vec a_2-\vec a_1
\, ,
\label{non-homo}
\ea
from which we extract
\ba
{\sf a}_R= {\sf b}\,,~~~~~~+_R= - \,,\nn \\
{\sf b}_R= {\sf a}\,,~~~~~~-_R= +\,,
\ea
and
\ba
{\vec t_{\sf a}^{\,R}}&=&-  \vec a_1\,,~~~~~~
\vec t_{\sf b}^{\,R}=\vec a_2-\vec a_1\,.
\ea
%
Second, for the $F_x$ generator we get
\ba
F_x \cdot \vec r_{\sf a} &=& \vec r_{\sf b}
- \vec a_2\,,
\nn\\
F_x\cdot \vec r_{\sf b} &=& \vec r_{\sf a}+
\vec a_1-\vec a_2
\, .
\label{non-homo2}
\ea
implying
\ba
{\sf a}_{F_x}= {\sf b}\,,~~~~~~+_{F_x}= +\,, \nn \\
{\sf b}_{F_x}= {\sf a}\,,~~~~~~-_{F_x}= -\,,
\ea
and
\ba
{\vec t_{\sf a}^{\,F_x}}&=&-  \vec a_2\,,~~~~~~
\vec t_{\sf b}^{\,F_x}= \vec a_1-\vec a_2\,.
\ea
In consequence  from \eqref{lachot} and \eqref{psi} we obtain
\be
D^{\bf R}(R)=\left(
            \begin{array}{cccc}
              &&0&\omega   \\
              &&\omega^*&0\\
              0&\omega^* \\
              \omega&0
            \end{array}
          \right)\,,
\label{dr}
\ee
\be
D^{\bf R}(F_x)=\left(
           \begin{array}{cccc}
              0&\omega^* &   \\
              \omega&0\\
              &&0&\omega \\
              &&\omega^*&0
            \end{array}
          \right)\,,
          \label{df}
\ee
with $\omega=e^{i\frac{2\pi}3}$.

The character for the $\sf D_6$ representation defined  by
\eqref{dr}-\eqref{df} is
\be
\chi^{\bf R}=(4,0,0,-2,0,0)\,.
\ee
This representation is  reducible and
decomposes in $\sf D_6$ irreducible representations as \cite{basko}
\be
\Psi_{ma} \in {\bf R}=\bf V\oplus V'\,.
\label{elbasko}
\ee

~

\noindent \ul{\sf Quadratic terms:}

~

Having obtained the representations of $\sf D_6$ contained on the basic low energy fields $u_{ij}(\vec x)$, $\partial_i h(\vec x)$
and $\Psi_{ma}(\vec x)$, our aim now is to build an invariant effective action.

As a first step, one might wonder whether quadratic fermionic terms, not including derivatives, are allowed in the low energy effective action. This amounts to
analyze the $\sf D_6$ singlets in the $ \Psi_{ma}^\dagger\Psi_{\ul{m}\ul{a}} $ decomposition.
From \eqref{CG} one immediately finds that the 16-dimensional fermion bilinear representation
decomposes as
\be
 \Psi_{ma}^\dagger\Psi_{\ul{m}\ul{a}} \in {\bf R}^*\otimes {\bf R} =2 {\bf E} \oplus 2 {\bf A}_1 \oplus 2 {\bf A}_2 \oplus 2 {\bf A}_3\oplus 2 {\bf V} \oplus 2 \bf V'\,.
\label{decompPSIPSI}
\ee
The coefficient of the singlet $\bf E$ in equation \eqref{decompPSIPSI} implies
that two possible memory tensors $M,\tilde M$ exist. Upon computation one finds
\ba
M&=&{\mathbb I}_{4\times 4}
\nn\\
\tilde M &=& \left(
\begin{array}{cccc}
&&0&1\\
&&1&0\\
0&1&&\\
1&0&&
\end{array}
\right)
\label{M}
\ea
The appearance of the $M$ tensor  is not surprising, since  the fermion representation defined by \eqref{dr}-\eqref{df} is easily seen to be unitary. By looking at the structure of the $\tilde M$ tensor, one realizes that it leads to fermion bilinears $\Psi^\dagger \tilde M \Psi$  that mix both the Fermi points $\pm$ and  the sublattices ${\sf a},{\sf b}$. As discussed at the end of Section \ref{methodology}, conservation of crystalline momentum does not allow for this last kind of terms in the Lagrangian, thus we must not consider $\tilde M$ in the low energy limit.
We conclude that, at the quadratic order and with no derivatives, symmetry allows only one term in the Lagrangian constructed  from the trivial tensor $M$\footnote{In order to have gap opening terms, it is necessary to relax some of the assumed symmetries \cite{semenoff,haldane,hou,chayu}}, which reads
\be
{\cal L}_\mu=\mu\, \Psi^\dagger \Psi \,.
\label{mass}
\ee
(here and in what follows the $ma$  indices are not made explicit whenever they are summed over). The arbitrary constant $\mu$ can be identified with a chemical potential.

The tensor $M$ also allows to write a kinetic term of the form
\be
{\cal L}_{kin}=  i\Psi^\dagger\partial_t\Psi\,.
\label{kinkin}
\ee

Next, one can explore the possible derivative terms. To do that, we have to decompose the product $\Psi^\dagger_{ma} \partial_i \Psi_{\ul m\ul a}$ into irreducible representations. This gives
\be
\Psi_{\!ma}^\dagger \partial_i \Psi_{\!\ul m \ul a}\! \in \!{\bf V} \otimes {\bf R}^*\!\otimes {\bf R} \!=\!
2 {\bf E} \oplus 2 {\bf A}_1 \oplus 2 {\bf A}_2 \oplus 2 {\bf A}_3\oplus 6 {\bf V} \oplus 6 \bf V'\!.
\label{decompPSIDPSI}
\ee
The presence of two singlets in the above decomposition,  whose origin can be traced back to the
presence of two $\bf V$ terms in \eqref{decompPSIPSI}, can be
understood as the existence of two independent sets of invariant ``Dirac gamma'' matrices $\gamma_i$ and $\tilde\gamma_i$
satisfying
\be
(D^{\bf R} (P))^\dagger\gamma_i\,D^{\bf R} (P)=D^{\bf V}_{ij}(P)\,\gamma_j\,.
\label{gammatransf}
\ee
In other words, the bilinears $\Psi^\dagger\gamma_i\Psi$ and $\Psi^\dagger\tilde\gamma_i\Psi$ constructed out from those $\gamma$-matrices, transform as vectors. As for the tensors derived above \eqref{M}, due to momentum conservation only one of the sets satisfying \eqref{gammatransf} is permitted in the low energy effective action. Indeed, upon analyzing the structure of the gamma matrices $\gamma,\tilde\gamma$
one observes that the fermion bilinear that results from the $\gamma$ set preserves the Fermi points while the one that results from the $\tilde\gamma$ set mixes the two Fermi points. Again, by crystalline momentum conservation we must discard the $\tilde \gamma_i$ set in the low energy effective action.
The relevant set  of Hermitian $\gamma$-matrices satisfying \eqref{gammatransf} reads
\ba
\gamma_1&=& \left(
    \begin{array}{cccc}
      &e^{-i2\pi/3} && \\
      e^{i2\pi/3}&& \\
      &&&e^{-i\pi/3}  \\
      &&e^{i\pi/3}
    \end{array}
  \right)\\
\gamma_2&=&\left(
    \begin{array}{cccc}
      &e^{-i\pi/6} && \\
      e^{i\pi/6}&& \\
      &&&e^{-i5\pi/6}  \\
      &&e^{i5\pi/6}
    \end{array}
  \right)
\label{gammas}
\ea
The reason for calling these tensors $\gamma$-matrices follows from their anti-commutator algebra
\be
\{\gamma_i,\gamma_j\}=2\delta_{ij}
\label{clifford}
\ee
and from the form of the singlets we can construct out of them.

 From all these ingredients it now follows that the
$\sf D_6$-invariant term that can  be constructed out of fermion bilinears and one derivative is
\be
{\cal L}_{v }= iv \sum_{ i}\Psi^\dagger\gamma_i\partial_i\Psi\,,
\label{kinfin}
\ee
where $v$ is an arbitrary constant. As shown below, this expression coincides with the derivative term originally obtained in the
literature from a tight-binding approach. The virtue of our approach is that it makes manifest
the transformation properties of the fermionic excitations under the point group.

~

\noindent \ul{\sf Accidental rotation symmetry}:

~

From the fermionic terms we just constructed \eqref{mass},\eqref{kinkin} and \eqref{kinfin}, we conclude   that, at the quadratic level in fermions, the  Lagrangian for graphene reads
\be
{\cal L}_2=i\Psi^\dagger\partial_t\Psi+iv \sum_{ i}\Psi^\dagger\gamma_i\partial_i\Psi+\mu\Psi^\dagger\Psi
\label{eledos}
\ee
This Lagrangian can be shown to have an accidental {\em continuous} symmetry, already discussed in \cite{basko} where it was called
``full intravalley rotational symmetry''. In our formalism, this symmetry arises naturally from the fermionic rotation generator $\Lambda$, constructed from the $\gamma$  as
\be
\Lambda=\frac i4\left[\gamma_1,\gamma_2\right]=\frac12\left(
          \begin{array}{cccc}
                    1   \\
                     & -1 \\
                     &&-1\\
                     &&&1
                  \end{array}
                \right) \,.
\ee
 From it  we can define the continuous matrix $D(\theta)$ as
\ba
D(\theta)&=&e^{i\theta \Lambda}=\left(
                                                          \begin{array}{cccc}
                                                            e^{i\frac\theta2}   \\
                                                            & e^{-i\frac\theta2}  \\
                                                            && e^{-i\frac\theta2}  \\
                                                            &&& e^{i\frac\theta2}  \\
                                                          \end{array}
                                                        \right).
\label{rot}
\ea
  which satisfies the relation
\ba
(D(\theta))^\dagger\gamma_i\,D(\theta)&=&R_{ij}(\theta)\,\gamma_j\,,
\label{fermrotation}
\ea
with
\ba
~~~~~~~~R(\theta)&=&\left(
          \begin{array}{cc}
                    \cos\theta & -\sin\theta \\
                    \sin\theta & \cos\theta \\
                  \end{array}
                \right)\,.
\ea
 Expression \eqref{fermrotation}  ensures that the quadratic Lagrangian ${\cal L}_{2}$ is invariant under the simultaneous transformations
\ba
\vec x\,'&=& R(\theta) \cdot \vec x \,,
\nn\\~
\nn\\
\Psi'_{ma}(\vec x)&=& \sum_{\ul{ma}}[D(\theta)]_{ma \ul{ma}}\Psi_{\ul{ma}}({(R(\theta))}^{-1}\!\!\cdot\vec x)\,.
\label{ferrot}
\ea
Notice that  upper and lower components of $\Psi_{ma}(\vec x)$ do not mix under \eqref{rot}.

Being an accidental symmetry of the lower derivative quadratic action, this symmetry does not need to be preserved by interactions nor by higher derivative terms, and as we will see in the forthcoming sections, it is indeed broken by electron-phonon interactions.

~

~

\noindent \ul{\sf Electron-phonon couplings}:

~

To construct the invariant terms coupling phonons to electrons, let us first concentrate in the in-plane phonons $u_{ij}(\vec x)$, which transforms in the ${\cal S}({\bf V}\otimes {\bf V})$ representation. We hence need to look for singlets in the decomposition of the product
\ba
u_{ij}  \Psi_{ma}^\dagger\Psi_{\ul{ma}} \in &&\!\!\!\!\!\!
{\cal S}({\bf V}\oplus{\bf V}')\otimes {\bf R}^*\otimes {\bf R}=\label{decompDUPSIPSI}
\\
 &=& 4 {\bf E} \oplus 4 {\bf A}_1 \oplus 4 {\bf A}_2 \oplus 4 {\bf A}_3\oplus 8 {\bf V} \oplus 8 \bf V'\, , \nn
\ea
In the first line of this equation, we can use \eqref{pho} and \eqref{decompPSIPSI} to advance that we will have four  singlets in the second line: two of them coming from the singlet in \eqref{pho} multiplied by the two singlets in \eqref{decompPSIPSI}, and the other two coming from the ${\bf V}'$ in $\eqref{pho}$ multiplied by the two ${\bf V}'$ in \eqref{decompPSIPSI}. Similarly to what happened with the quadratic invariants, momentum conservation precludes half of them, therefore in what follows we only write down the two tensors which do not mix the Fermi points. The one coming from the product of the singlet in \eqref{pho} with the singlet in \eqref{decompPSIPSI} reads
\be
{\cal L}_{\bar q_{(u)} }=\bar q_{(u)}\, \Psi^\dagger\Psi \sum_{i}  u_{ii} \,.
\label{E}
\ee
This term has its physical origin in the changes in the area of the unit cell due to contractions and dilatations\footnote{See eqs. (150) and (169) in \cite{review}}.

  The singlet arising from the product of the  ${\bf V}'$ in \eqref{pho} with the ${\bf V}'$ in \eqref{decompPSIPSI}, in complete analogy with the discussion above on the derivative terms, implies the existence of a set of gamma matrices $\bar\gamma_{\alpha}$ that project the fermion bilinear \eqref{decompPSIPSI} onto its relevant $\bf V'$ component. They read
\ba
\bar\gamma_1&=& \left(
    \begin{array}{cccc}
      &e^{i\pi/3} && \\
      e^{-i\pi/3}&& \\
      &&&e^{-i\pi/3}  \\
      &&e^{i\pi/3}
    \end{array}
  \right)\\
\bar\gamma_2&=&\left(
    \begin{array}{cccc}
      &e^{-i\pi/6} && \\
      e^{i\pi/6}&& \\
      &&&e^{i\pi/6}  \\
      &&e^{-i\pi/6}
    \end{array}
  \right)
\label{gammasss}
\ea
and satisfy
\be
(D(P))^\dagger\,\bar\gamma_\alpha\,D(P)=D^{\bf V'}_{\alpha\beta} (P)\,\bar\gamma_\beta\,.
\ee
The resulting interaction term has the form
\be
{\cal L}_{q_{(u)}}=q_{(u)}\sum_{\alpha}\Psi^\dagger \bar\gamma_\alpha\Psi \,A^{(u)}_\alpha\,,
\label{v'}
\ee
where the pseudovector $A^{(u)}_\alpha$ was obtained in
\eqref{Adu} by projecting the phononic components into the ${\bf V}'$ representation making use of the tensor $S_{\alpha ij}$ defined in \eqref{nicoviri}. This interaction term reproduces the expressions previously obtained in the literature \cite{review}.

~


Now we can turn into the invariant terms coupling flexural phonons to electrons. To lowest degree in fields and lowest orders in derivatives, we need to look for singlets in the decomposition of the product
\be
\partial_i h \Psi_{\!ma}^\dagger \Psi_{\!\ul m \ul a}\! \in \!{\bf V} \otimes {\bf R}^*\!\otimes {\bf R} \,.
\label{decompPSIDhPSI}
\ee
Since the irreducible representations entering into this product are exactly the same as in \eqref{decompPSIDPSI}, we repeat the analysis we made to write the quadratic derivative terms and obtain the invariant coupling
\be
{\cal L}_{g}=v g \sum_{i}\Psi^\dagger \gamma_i\Psi  \,\partial_i h \, .
\label{trcoup}
\ee

To the same order in the fields, we can easily go to higher orders in derivatives. Since $\partial_i\partial_jh$ transforms in the same representation as $u_{ij}$, eq. \eqref{decompDUPSIPSI} and the subsequent considerations immediately tell us that the allowed invariant couplings are
\be
{\cal L}_{\bar q_{(h)}}=\bar q_{(h)}\Psi^\dagger\Psi\nabla^2h\,,
\ee
and
\be
{\cal L}_{q_{(h)}}=q_{(h)}\sum_{\alpha}\Psi^\dagger \bar\gamma_\alpha\Psi \,A^{(h)}_\alpha\,,
\label{vc}
\ee
where
\ba
A^{(h)}_\alpha(\vec x)&=&\sum_{ij}S_{\alpha ij}\,\partial_i\partial_jh(\vec x)\,.
\ea
The interaction term \eqref{vc} has been recently shown to appear when a coupling to the spin degrees of freedom is present \cite{cngui}.

Notice that in the absence of a substrate, the transformation $h(\vec x)\to -h(\vec x)$ must be a symmetry of the Lagrangian, which implies that the coupling constants $g,\bar q_{(h)}$ and $q_{(h)}$ must be set to zero. In that case, the first non-trivial interaction arises to quadratic order in $\partial_ih(\vec x)$. Again since $\partial_ih(\vec x)\partial_jh(\vec x)$ transform in the same representation as $u_{ij}(\vec x)$, we can write down the relevant terms without any additional analysis. They read
\be
{\cal L}_{\bar q_{(hh)}}=\bar q_{(hh)}\Psi^\dagger\Psi(\nabla h)^2\,,
\ee
and
\be
{\cal L}_{ q_{(hh)}}= q_{(hh)}\sum_{\alpha}\Psi^\dagger\bar \gamma_\alpha\Psi \,A^{(hh)}_\alpha\,,
\label{vculo}
\ee
where
\ba
A^{(hh)}_\alpha(\vec x)&=&\sum_{ij}S_{\alpha ij}\,\partial_ih(\vec x)\partial_jh(\vec x)\,.
\ea

The procedure just presented can be pursued to any degree in the fields and to any order in derivatives, allowing for the construction of the most general effective action for electrons coupled to phonons in graphene.
%
%
%
%
%
%
%

~

\noindent \ul{\sf Final result and comparison with the literature}:

~

Then, as a general conclusion, to the lowest order in the fields and their derivatives, the most general Lagrangian consistent with the symmetries read
\ba
{\cal L}\!&=&\!i \Psi^\dagger \partial_t \Psi +i v\sum_i\Psi^\dagger \gamma_i\left(\partial_i -ig\partial_ih\right)\Psi
+\nn\\
&&
+\!
\left(\mu+
\bar q_{(u)}\sum_iu_{ii}
+
\bar q_{(h)}\nabla^2h
+
\bar q_{(hh)}(\nabla h)^2
\right)\Psi^\dagger\Psi+
\nn\\&&
+
\sum_{\alpha}
\left(
q_{(u)} A^{(u)}_\alpha+
q_{(h)} A^{(h)}_\alpha+
q_{(hh)} A^{(hh)}_\alpha
\right)
\Psi^\dagger \bar\gamma_\alpha\Psi \,.\nn\\
\label{full}
\ea
The first observation regarding this Lagrangian is that even if the first two lines in \eqref{full} are invariant under the accidental continuous symmetry \eqref{ferrot}, the third line is not. This is due to the fact that the tensor $S_{\alpha ij}$ entering into the construction of the pseudovector fields $A_\alpha(\vec x)$ is not invariant under $R(\theta)$ but only under transformations in $G_P$. A second observation is that the coupling to $\partial_ih(\vec x)$ in the first line can be reabsorbed by a local phase redefinition of the electronic fields $\Psi'_{ma}(\vec x)=\exp(-igh(\vec x))\Psi_{ma}(\vec x)$. Nevertheless, such operation would give rise to a kinetic coupling $-g\Psi^\dagger\Psi \,\partial_t h$. A final point is that, since the $A_\alpha(\vec x)$ fields are pseudovectors, a local phase redefinition of $\Psi_{ma}(\vec x)$ cannot be reabsorbed by a derivative shift in any of them. In other words, the fields $A_\alpha(\vec x)$ are not ``gauge'' fields in the standard sense.

~

We close this section showing the consistency of our results with the existing effective
Lagrangians computed from a tight-binding approach, to lowest order in momentum and deformation fields.
In order to make contact with the literature, we rewrite the electron field as
%
%
\be
\psi_+ = \left(
\begin{array}{c}
e^{-i\frac\pi3}\Psi_{+{\sf b}} \\
e^{i\frac\pi3}~\Psi_{+{\sf a}}
\end{array}
\right)\qquad
\psi_- = \left(
\begin{array}{c}
e^{-i\frac\pi6}\Psi_{-{\sf b}} \\
e^{i\frac\pi6}~\Psi_{-{\sf a}}
\end{array}
\right)\,.
\ee

From our kinetic Lagrangian ${\cal L}_{v }$ in \eqref{kinfin} we can extract the quadratic part of the Hamiltonian density
\ba
H_{{v}}
&=&-iv \sum_{i}\left(
\psi_+^{\dagger}\,\sigma_i\partial_i\psi_+ +\psi_-^{\dagger}\,\bar\sigma_i\partial_i\psi_-
\right)\,,
\label{kinnos}
\ea
where $\sigma_i=(\sigma_1,\sigma_2)$ are the usual Pauli matrices,
$\bar\sigma_i=(\sigma_1,-\sigma_2)$. This coincides with the result obtained in \cite{AS}.

As for the electron-phonon coupling, the term arising from \eqref{E} reads
\be
H_{\bar q_{(u)}}= \bar q_{(u)}\, \left(
\psi_+^\dagger\psi_+
+
\psi_-^\dagger\psi_-
\right)
\sum_{i}u_{ii}\,,
\ee
and coincides exactly with the one presented in ref. \cite{AS}.
\
Finally from \eqref{v'} one obtains
\ba
H_{q_{(u)}}
&=&-q_{(u)}\sum_{\alpha} \left(
\psi_+^{\dagger} \bar\sigma_\alpha  \psi_{+}- \psi_-^{\dagger}\sigma_\alpha \psi_{-}\right)
\, A_\alpha^{(u)}\,,
\label{nosint}
\ea
where $\sigma_\alpha =(\sigma_1, \sigma_2)$ and $\bar\sigma_\alpha =(\sigma_1, -\sigma_2)$. This again coincides with eq. (3.9) of \cite{AS}.
%
%
%
%
%
%
%

~

\paragraph{The kagom\'e lattice case}

~

~

We now sketch the application of our technique to the kagom\'e lattice, highlighting
the main differences with the graphene case discussed above.

The kagom\'e lattice can be seen as a triangular lattice with three atoms per
fundamental cell. As for the graphene case we consider the simplest non-trivial
case of a doubly degenerate low energy band structure with two inequivalent Dirac points,
$\sf D_6$ symmetry  implies that they should be located at the vertices $\vec k_\pm$ of the Brillouin zone. The third band is not degenerate and it will be considered a high energy band. This instance
in particular includes the description of the kagom\'e lattice at 1/3 filling. In consequence, we have six degrees of freedom at low energies, that we arrange as
\be
\Psi_{ma}=
\left(
\begin{array}{c}
\Psi_{+{\sf a}}\\
\Psi_{+{\sf b}}\\
\Psi_{+{\sf c}}\\
\Psi_{-{\sf a}}\\
\Psi_{-{\sf b}}\\
\Psi_{-{\sf c}}
\end{array}
\right)\,,
\ee
%

To derive the transformation properties of the six degrees of freedom, we consider the
triangular lattice to be generated by \eqref{bravaisvect}. The position of the $a={\sf a},{\sf b},{\sf c}$ atoms inside the unit cell are
\be
\vec r_{\sf a}=\frac1{2}(\vec a_1+\vec a_2),~~~~\vec r_{\sf b}=\frac12\vec a_2,~~~~\vec r_{\sf c}=\frac12\vec a_1.
\ee
The action of the {\sf D}$_6$ generators $R,F_x$ on the basis $\vec r_a$ is
\ba
R \cdot\vec r_{\sf a}&=&\vec r_{\sf c}+\vec a_2-\vec a_1\,,
\nn\\
R\cdot \vec r_{\sf b}&=&\vec r_{\sf a}-\vec a_1\,,
\nn\\
R\cdot \vec r_{\sf c}&=&\vec r_{\sf b}\,.
\label{non-homokagomeR}
\ea
and
\ba
F_x \cdot\vec r_{\sf a}&=&\vec r_{\sf b}+\vec a_1-\vec a_2\,,
\nn\\
F_x\cdot \vec r_{\sf b}&=&\vec r_{\sf a}-\vec a_2\,,
\nn\\
F_x\cdot \vec r_{\sf c}&=&\vec r_{\sf c}\,,
\label{non-homokagomeF}
\ea
from where one reads the information needed for computing \eqref{lachot}, as
\ba
{\sf a}_R= {\sf c}\,,~~~~~~+_R= - \,,\nn \\
{\sf b}_R= {\sf a}\,,~~~~~~-_R= + \,,\nn \\
{\sf c}_R= {\sf b}\,.~~~~~~\phantom{-_R= + \ \, ,}
\ea
and
\ba
{\vec t_{\sf a}^{\,R}}&=&\vec a_2-\vec a_1\,,~~~~~~
\vec t_{\sf b}^{\,R}=-\vec a_1\,,~~~~~~
\vec t_{\sf c}^{\,R}=0\,,
\ea
for the $R$ generator, while for the $F_x$ generator we get
\ba
{\sf a}_{F_x}= {\sf b}\,,~~~~~~+_{F_x}= - \,,\nn \\
{\sf b}_{F_x}= {\sf a}\,,~~~~~~-_{F_x}= + \,,\nn \\
{\sf c}_{F_x}= {\sf c}\,,~~~~~~\phantom{-_{F_x}= +\; \,, }
\ea
and
\ba
{\vec t_{\sf a}^{\,F_x}}&=&\vec a_1-\vec a_2\,,~~~~~~
\vec t_{\sf b}^{\,F_x}=-\vec a_2\,,~~~~~~
\vec t_{\sf c}^{\,F_x}=0\,.
\ea
In consequence we obtain
\be
D(R)=\left(
            \begin{array}{cccccc}
              & & & 0& \omega^*&0 \\
              & & &  0& 0&1\\
              & & & \omega& 0&0 \\
              0&\omega & 0& & & \\
              0&0 &1 & & & \\
              \omega^*&0 &0 & & &
            \end{array}
          \right)\,,
\label{non-homokagR}
\ee
and
\be
D(F_x)=\left(
            \begin{array}{cccccc}
              0&\omega &0 & & & \\
              \omega^*&0 &0 & & &\\
              0&0 & 1& & & \\
              &&&0&\omega^* &0 \\
              &&&\omega&0 & 0 \\
              &&&0& 0&1
            \end{array}
          \right)\,.
\label{non-homokagF}
\ee
%
%
%
%
The character for the representation defined  by
\eqref{non-homokagR}-\eqref{non-homokagF} is
\be
\chi^{\bf R}=(6,0,0,0,2,0)\,.
\ee
Upon decomposing it, one finds
\be
\Psi_{ma} \in {\bf R} = {\bf E\oplus A}_3\oplus\bf V\oplus V' .
\label{psikagome}
\ee
Two additional degrees of freedom appear in this decomposition when compared to the case of graphene ({\em i.e.} the scalar ${\bf E}$ and the pseudoscalar ${\bf A}_3$).   As we explain below, they give rise to high energy bands and are naturally projected out for generic values of the couplings in the resulting effective Lagrangian.

~

\noindent \ul{\sf Quadratic terms}:

~

We now turn to the construction of the possible low energy effective terms in the Lagrangian.
Having defined the transformation properties of $\Psi_{ma}$ under $\sf D_6$, these terms are obtained from
the projection of the appropriate field products into $\sf D_6$ singlets. As discussed for the graphene case,
crystalline momentum conservation is equivalent to demanding that the terms entering into the Lagrangian must not mix the Dirac points.

The decomposition of a fermion bilinear results in
\be
 \Psi_{ma}^\dagger\Psi_{\ul{ma}} \in {\bf R}\otimes {\bf R}=4 {\bf E} \oplus 2 {\bf A}_1 \oplus 2 {\bf A}_2 \oplus 4 {\bf A}_3\oplus 6 {\bf V} \oplus 6 \bf V'\,.
\label{decompPSIPSIKagome}
\ee
Following the same steps as in the case of graphene, the resulting effective Lagrangian to quadratic order in the fields reads
\be
{\cal L}_{2} = \mu \, \Psi^\dagger\Psi
+\Delta\,\Psi^\dagger \tilde M\Psi
+i
\sum_{i} \Psi^\dagger(
\check v\,\check\gamma_i+\tilde v\,\tilde\gamma_i+ \hat v\,\hat\gamma_i)\partial_i\Psi\,,
\label{effham}
\ee
where
\be
\tilde M=\left(
    \begin{array}{cccccc}
      0 & \omega & \omega^*&  \\
      \omega^* & 0 & \omega \\
      \omega & \omega^* & 0 \\
      &&& 0& \omega^* & \omega\\
      &&& \omega& 0 & \omega^*\\
      &&& \omega^*& \omega & 0
    \end{array}
  \right)\,,
\ee
and the explicit expression of the three sets of $\gamma$-matrices is given in appendix \ref{apgamma}.

The coefficient $\mu$ in \eqref{effham} amounts to a chemical potential for the electrons.
The interesting new ingredient in the Kagome lattice is the $\tilde M$ invariant tensor.
In the abasence of $\tilde M$ ($\Delta=0$) we obtain, at zero momentum, six degenerate states. The existence
of a non-zero $\Delta$ allows for a splitting among this $6$ states into $4+2$ degenerate states. This fact
 leads to the interpretation of the coefficient $\Delta$ as setting a gap between the six states contained in $\Psi$.
Therefore the degeneracy of the lowest energy state at zero momentum can   be set to 6, 2 or 4 depending on
whether $\Delta=0,\Delta>0$ or $\Delta<0$. The lowest order correction in momentum to the ground state is non-zero  for the four degenerate modes (which get split into 2+2), while to linear order the two splitted states get no correction.

We can now see a beautiful agreement of our group theory approach with the phenomenology of the Kagom\'e system:
 a tight binding approach applied to fermions hopping to nearest neighbors in the kagom\'e lattice \cite{Kagome,hyper}
shows that at filling $1/3$  only four degrees of freedom appear at low energy\footnote{The dispersion relation for
the three bands found in \cite{Kagome,hyper} show two bands similar to the ones found in graphene and a third one
being flat and separated from the first two  by a gap set by the hopping parameter.}. The  existence  of
$\tilde M$ from the group theory approach allows a nice match with this result if we set $\Delta$ to be positive and very large. Notice that in that case
the assumption that the high-energy bands can be expanded around the minima of the low-energy bands is innocuous.

At this point, a consistent low energy expansion demands to project out the massive modes.
Implementing this projection amounts to eliminate the irreducible representations $\bf E$ and
${\bf A}_3$ from \eqref{psikagome}\footnote{One can check that the two degenerate modes of $\tilde M$ correspond to
the $\bf E$ and ${\bf A}_3$ inside $\Psi$.}, which then implies that the low energy effective action coincides with
the one found in the graphene case.

~

\noindent \ul{\sf Electron-phonon coupling}:

~

From this simple analysis we conclude that at low energy, the electron-phonon coupling terms
for the kagom\'e lattice have the same expressions as in the graphene case,
without the need of going through detailed tight-binding computations.
More explicitely, for large $\Delta$ \eqref{psikagome} matches exactly \eqref{elbasko} and hence the decomposition \eqref{decompDUPSIPSI} would be the same, implying that the interaction terms would be given by \eqref{E}, \eqref{v'}.

This result shows the
power of our symmetry based approach to derive low energy effective actions.

\section{conclusion}
\label{conclusions}
In the present paper we provide a method to construct the effective low energy Lagrangian of an electronic system defined on a given lattice. We use a symmetry based approach that incorporates the notion of memory tensors, that are nothing but the invariant tensors of the point group. More specifically, we have revisited in detail how symmetries constrain the low energy dynamics of a generic lattice system and how to retain the necessary specific information about the underlying discrete symmetries.

The main idea is that in the infrared, the discrete translations of the lattice become continuous translations, while the point group retains its discrete nature. In other words, the low energy dynamics is described by a field theory with fields transforming in representations of the discrete point group $G_P$. We have shown how to construct systematically the infrared effective Hamiltonian, invariant under the full symmetry group, encoding the discrete origin of the system into the invariant tensors. In other words, the low energy dynamics of a system of electrons in a given lattice would be determined by the content of memory tensors of $G_P$ and that allows to construct all the different invariant local terms in the effective Hamiltonian.

We have applied the method developed to the case of the $\sf D_6$ point group, and focused into both graphene and kagom\'e systems. For the graphene case we have shown how to derive the low energy effective field theory reproducing the tight-binding expressions for the electron-phonon system \cite{AS,KCN}. In a second step, we have applied the method to electrons in the kagom\'e lattice close to filling $1/3$ \cite{Kagome,hyper}. In this example, the method shows its power since the derivation of the effective action is quite straightforward.

The advantages of the present approach is that it can be applied to study more general cases, {\it i.e.} lattices with other point groups, higher order in the fields and/or their derivatives, in a straightforward manner.

Using the symmetry based approach presented in this paper one could envisage the study of situations in which lattice deformations contribute to next to leading orders in a systematic way.  We believe that, for the particular case of two-dimensional
 systems, as graphene, it is possible to overcome the small deformations expansion. These ideas will be explored in future investigations. This
could be a particularly important step forward in modeling some experimental setups in which deformation of graphene sheets goes beyond small perturbations of the flat lattice. Another possible generalization that can be easily envisaged within the present symmetry based set up is that of the inclusion of boundaries and point-like defects. Our procedure could also be applied to three dimensional systems, such as hyper-pyrochlore lattice studied in \cite{hyper}.

\vskip 1cm

~~

\noindent {\bf Acknowledgements:} We thank J.L. Ma\~nes for
correspondence. This work was partially supported PICT ANPCYT Grants N$^o$
20350 and 00849, PIP CONICET Grants N$^o$ 1691 and N$^o$ 0396, and Spanish MECD Grants N$^o$ FIS2011-23712 and N$^o$ PIB2010BZ-00512. M.S. thanks
M. Vozmediano, F. de Juan, A. Cortijo and A.G. Grushin for enlightening discussions. N.E.G. thanks A. Maharana and M. Torabian for helpful comments.

\newpage

\appendix

\section{An alternative approach for determining the fermion transformation properties}

In this appendix we show that an alternative approach could be used to derive the transformation
properties of the low energy degrees of freedom.

~

\noindent{\sf Graphene:} From the hypothesis of two inequivalent Dirac points, we
expect four degrees of freedom at low energy.
They correspond to:
(i) the two atoms inside the fundamental cell which we denote
$n=a,b$, and (ii) the two inequivalent Dirac points in Fourier space at which the Fermi surface degenerates,
which we denote $N=1,2$ (necessarily at the vertices of the BZ). With the additional input of a linear dispersion relation
for the low energy degrees of freedom, we will be able to deduce the same transformation properties as obtained in the text.

We start by constructing a linear representation of the $ \sf D_6$  point symmetry group acting on
these four degrees of freedom. To this end, we place them
inside a $4$-tuple $\Psi$ as\footnote{In what follows we will omit the momentum dependence of $\Psi$.}
\be
\Psi=\left(\begin{array}{c}
             a_1 \\
             b_2 \\
             a_2 \\
             b_1
           \end{array}
\right)
\label{LEexcitations}
\ee
To fix the transformation
properties of these degrees of freedom under $G_P$ we orient the hexagon so that two of its
vertices lay on the $y$-axis. When Fourier transforming, the first Brillouin zone
consists of an hexagon, now rotated $90^o$. Our conventions coincide with  those in \cite{Manes07},\cite{Winkler}.

It is immediate
to see that a fundamental rotation $R$ interchanges both the atoms and the Dirac points,
\be
a\leftrightarrow b,~~~~~1\leftrightarrow2
\ee
We therefore propose
\be
\begin{array}{c}
  R\,a_1=\alpha\, b_2\,, \\
  R\, b_2=\beta\, a_1\,,
\end{array}
\label{alfabeta1}
\ee
with phases $\alpha,\beta$ to be determined below from the point group constraints in \eqref{constraints}
and the linear dispersion relation conditions.

Time reversal
$\cal T$ symmetry fixes  the transformations of the remaining
degrees of freedom ($a_2,b_1$) as we now explain. First notice that in
the absence of spin, $\cal T$ simply acts as complex conjugation.
Taking into account the
dependence of the wave function on momentum
one concludes that time reversal interchanges the Dirac points and
changes the sign of the momentum with no action on the atoms positions (see \cite{Manes07} for a detailed discussion). One therefore
has
\be
\begin{array}{c}
  {\cal T}\,a_1=a_2 \\
  {\cal T}\, b_1= b_2\,.
\end{array}
\ee
Acting with $\cal T$ on \eqref{alfabeta1}, one obtains
\be
\begin{array}{c}
  R\,a_2=\alpha^*\, b_1 \\
  R\, b_1=\beta^*\, a_2\,.
\end{array}
\label{alfabeta2}
\ee
The phases $\alpha,\beta$ are not completely arbitrary,  we should demand that the transformation $D(R)$ defined by
\eqref{alfabeta1}-\eqref{alfabeta2} satisfies the group constraints \eqref{constraints}, that is
\be
(D(R))^6\Psi=\Psi~~~\Rightarrow~~~(\alpha\beta)^3=1
\label{graphcons}
\ee
Let us now analyze the action of a $x$-axis reflection on the fermions. For the cell orientation
we have chosen, it is easy to see that it preserves the Dirac
points but interchanges the atoms,
\ba
  F_x\,a_1&=& b_1 \\
  F_x\, b_1&=&  a_1\,.
  \label{fx1}
\ea
The transformation of $(a_2,b_2)$ is again fixed by time reversal to be
\ba
  F_x\,a_2&=& b_2 \\
  F_x\, b_2&=&  a_2\,.
  \label{fx2}
\ea
The definitions \eqref{fx1}-\eqref{fx2} automatically satisfy the second constraint in \eqref{constraints}
\be
(D(F_x))^2\Psi=\Psi\,.
\ee
The remaining non-trivial constraint among the $\sf D_6$ generators
\be
F_x R F_x=R^5
\label{cons2}
\ee
gives no new constraint on the phases
other than the one we have already found in \eqref{graphcons}.

From \eqref{graphcons} one finds three possibilities for representing $\sf D_6$ on $\Psi$,
\be
\alpha\beta=1,
\label{nochoice}
\ee
or
\be
\alpha\beta=e^{\pm i\frac{2\pi}3}
\label{possibles}
\ee
These last two possibilities can be shown to be equivalent
upon computing the character of the representation\footnote{The character for $D$ is $\chi=(4,0,0,2(\alpha\beta+\alpha^*\beta^*),0,0)$).}.

Now the requisite of a linear dispersion relation at low energy for the fermions comes into play.
This condition requires an effective action containing a $\sf D_6$ singlet constructed out of two
fermions and one spatial derivative. Since the derivative transforms in the representation $\bf V$, \eqref{CG} implies
that the only possibility for a singlet comes from  contracting $\partial$ with another $\bf V$ representation.
The second choice in \eqref{possibles} guarantees that this will be possible (in principle two possible singlets can be obtained when combining
two fermions  with a derivative see the text)\footnote{We realize the appropriate choice \eqref{possibles} as
\be
\alpha=\beta=\omega=e^{i\frac{\pi}3}\,.
\label{choice}
\ee}. Hence, {\it the fermion dispersion relation at low energy dictates the choice of phases
for the representation.}\footnote{A fortiori, one can check that the representation \eqref{irr0},\eqref{irr}
is equivalent to that proposed in \cite{Manes07}. Had we chosen the first possibility
in \eqref{nochoice} the decomposition of $\Psi$ would have been
\be
\Psi\in{\bf E} \oplus  {\bf A}_1 \oplus  {\bf A}_2 \oplus  {\bf A}_3\,.
\ee
which does not allow for a standard kinetic term for the fermions in the sense explained above} In a generic case, a linear dispersion relation for fermions would be forbidden by $\sf D_6$ symmetry,
if no possible choice of phases could lead to a $\bf V$ component in the fermions bilinear decomposition.

~

\noindent {\sf Summarizing}:
the fermionic degrees of freedom \eqref{LEexcitations} mix under the point symmetry group as
\be
\Psi'=D(P)\,\Psi,
\ee
with
\be
D(R)=\left(
            \begin{array}{cccc}
              0&\alpha &   \\
              \alpha&0\\
              &&0&\alpha^* \\
              &&\alpha^*&0
            \end{array}
          \right)
\label{irr0}
\ee
\be
D(F_x)=\left(
            \begin{array}{cccc}
              &&0&1    \\
              &&1&0\\
              0&1 \\
              1&0
            \end{array}
          \right)
          \label{irr}
\ee
where $\alpha=e^{i\frac\pi3}$ corresponds to a $60^o$ rotation.
The representation \eqref{irr0}-\eqref{irr} is unitary, reducible and
decomposes in $\sf D_6$ irreducible representations as \cite{basko} (cf. \eqref{elbasko})
\be
\Psi=\bf V\oplus V'
\label{elbasko2}
\ee

~

\noindent{\sf Kagom\'e}: We now apply the procedure above to the kagom\'e lattice.
The kagom\'e lattice has three atoms $a,b,c$ per
fundamental cell ($n=3$), and as for the graphene case we consider the simplest non-trivial
case of a (degenerate) band structure with two inequivalent Dirac points ($N=2$).
We place the six degrees of freedom inside a $6$-tuple as
\be
\Psi=\left(\begin{array}{c}
             a_1 \\
             c_1 \\
             b_1 \\
             a_2\\
             c_2\\
             b_2
           \end{array}
\right) \label{kagfermsap}
\ee
Under a fundamental rotation $R$ the fermions mix simultaneously as
\be
a\to b\to c\to a,~~~~~1\leftrightarrow2\,.
\ee
Allowing for possible phases, this amount to
\be
\begin{array}{cc}
  R\,a_1&=\alpha\, b_2 \\
  R\, b_2&=\beta\, c_1\\
  R\, c_1&=\gamma\, a_2\,,
\end{array}\stackrel{\cal T}{\to}
\begin{array}{cc}
  R\,a_2&=\alpha^*\, b_1 \\
  R\, b_1&=\beta^*\, c_2\\
  R\, c_2&=\gamma^*\, a_1\,,
\end{array}
\label{abc}
\ee
where as before the phases will be determined from the constraints \eqref{constraints}.
A reflection through the $x$-axis preserves the cones, leaves the $c$ degrees of freedom invariant
and permutes $a\leftrightarrow b$, that is
\ba
  F_x\,a_1&=& b_1 \\
  F_x\, b_1&=&  a_1\\
  F_x\, c_1&=&  c_1\,.
\ea
The remaining degrees of freedom transformation properties are fixed by time reversal and amount to
similar expressions changing $1\leftrightarrow2$.

The first two constraints in \eqref{constraints} are automatically satisfied, and the constraint \eqref{cons2}
results in $\alpha$ undetermined and $\beta\gamma=1$.
Computing the character for the representation one finds that it is in fact
independent of the product $\beta\gamma$. It is easily seen that this is due to the fact that all phases
can be eliminated by appropriate definitions of the degrees of freedom.
The simplest choice is $\alpha=\beta=\gamma=1$.

Upon computing the character one finds that the fermionic degrees of freedom \eqref{kagfermsap} provide a representation for $\sf D_6$ which
coincides with the one deduced from the transformation properties of the Wannier states \eqref{psikagome}.
\newpage
\section{Kagome gamma matrices}
\label{apgamma}

First set:
\small
\be
\check\gamma_1= \left(\begin{array}{cccccc}
  -1/4\\
  &-1/4\\
  &&1/2\\
  &&&1/4\\
  &&&&1/4\\
  &&&&&-1/2
   \end{array}
  \right)
\ee
\be
\check\gamma_2=
  \left(\begin{array}{cccccc}
  \sqrt3/4\\
  &-\sqrt3/4\\
  &&0\\
  &&&-\sqrt3/4\\
  &&&&\sqrt3/4\\
  &&&&&0
   \end{array}
  \right)
\ee

~

Second set:
\small
\be
\tilde\gamma_1= \left(\begin{array}{cccccc}
  0&-e^{-i\frac\pi3}&1\\
  -e^{-i\frac\pi3}&0&e^{-i\frac{2\pi}3}\\
  1&e^{i\frac{2\pi}3}&0\\
  &&&0&e^{i\frac\pi3}&e^{i\frac\pi3}\\
  &&&e^{-i\frac\pi3}&0&-1\\
  &&&e^{-i\frac\pi3}&-1&0
   \end{array}
  \right)
\ee
\be
\tilde\gamma_2=\left(\begin{array}{cccccc}
  0&e^{i\frac\pi6}&e^{i\frac\pi2}\\
  e^{-i\frac\pi6}&0&e^{i\frac{5\pi}6}\\
  e^{-i\frac\pi2}&e^{-i\frac{5\pi}6}&0\\
  &&&0&e^{-i\frac\pi6}&e^{i\frac{5\pi}6}\\
  &&&e^{i\frac\pi6}&0&e^{i\frac\pi2}\\
  &&&e^{-i\frac{5\pi}6}&e^{-i\frac\pi2}&0
   \end{array}
  \right)
\ee

~

\vspace{1cm}
\begin{widetext}
Third set:

~

\be
\hat\gamma_1= \left(\begin{array}{cccccc}
  0&\sqrt3e^{i\frac{2\pi}3}&e^{i\frac{\pi}2}\\
  \sqrt3e^{-i\frac{2\pi}3}&0&e^{-i\frac\pi6}\\
  e^{-i\frac{\pi}2}&e^{i\frac\pi6}&0\\
  &&&0&\sqrt3e^{i\frac\pi3}&e^{i\frac{5\pi}6}\\
  &&&\sqrt3e^{-i\frac{\pi}3}&0&e^{-i\frac\pi2}\\
  &&&e^{-i\frac{5\pi}6}&e^{i\frac\pi2}&0
   \end{array}
  \right)
\ee

\be
\hat\gamma_2=\left(\begin{array}{cccccc}
  0&-\frac12-\frac i{2\sqrt3}&1+\frac{2i}{\sqrt3}\\
  -\frac12+\frac i{2\sqrt3}&0&-\frac12+\frac{5i}{2\sqrt3}\\
  1-\frac{2i}{\sqrt3}&-\frac12-\frac{5i}{2\sqrt3}&0\\
  &&&0&-\frac12+\frac i{2\sqrt3}&-\frac12+\frac{5i}{2\sqrt3}\\
  &&&-\frac12-\frac i{2\sqrt3}&0&1+\frac{2i}{\sqrt3}\\
  &&&-\frac12-\frac{5i}{2\sqrt3}&1-\frac{2i}{\sqrt3}&0
   \end{array}
  \right)
\ee
\end{widetext}

\normalsize
\newpage

~

\end{document}